\documentclass[fleqn,11pt]{article}
\usepackage{epsfig}
\input{amssym.def}
\input{amssym.tex}
\author{ }
\date{ }
\makeatletter

\def\compoundrel#1\over#2{\mathpalette\compoundreL{{#1}\over{#2}}}
\def\compoundreL#1#2{\compoundREL#1#2}
\def\compoundREL#1#2\over#3{\mathrel
      {\vcenter{\hbox{$\m@th\buildrel{#1#2}\over{#1#3}$}}}}
\begin{document}
%

\vspace*{1.5cm}

\hfill\ {\Large\bf KEK Preprint 2019-1}

\vspace*{3mm}
\hfill\ {\Large\bf March 2019~~~~~~~~~~~~~}

\vspace*{3mm}
\hfill\ {\Large\bf H~~~~~~~~~~~~~~~~~~~~~~~~~~}

\vspace*{3.0cm}

\begin{center}
{\Huge\bf A 15-MW Proton Driver for}\\
\vspace*{0.6cm}
{\Huge\bf Neutrino Oscillation Experiments}\\
\vspace*{5.4cm}
{\LARGE R. Belusevic}\\
\vspace*{0.8cm}
{\large IPNS, {\em High Energy Accelerator Research Organization} (KEK)}\\
\vspace*{1.3mm}
{\large 1-1 {\em Oho, Tsukuba, Ibaraki} 305-0801, {\em Japan}} \\
\vspace*{1.3mm}
{\large r.belusevic@gmail.com}\\
\end{center}

\thispagestyle{empty}

\newpage

\tableofcontents
\addtocontents{toc}{\protect\vspace{1.3cm}}
\vspace*{5mm}
\noindent
{\large\bf References}


\newpage

\begin{center}
\begin{minipage}[t]{13.6cm}
{\bf Abstract\,:}\hspace*{3mm}
{To maximize the physics potential of future neutrino oscillation experiments, it
is proposed to build a 15-MW `proton driver' consisitng solely of a 3-GeV proton
injector linac (PI) and a 17-GeV superconducting ILC-type linac (SCL). The
proposed facility would provide proton beams with intensities more than an order
of magnitude higher than those at the existing proton synchrotron complexes.
Multi-MW proton beams could also be used to produce high-intensity antiproton, `cold' neutron, kaon, pion and muon beams for a diverse program of experiments in
particle and nuclear physics.}
\end{minipage}
\end{center}

\vspace*{0.3cm}
\renewcommand{\thesection}{\arabic{section}}
\section{Introduction}
\vspace*{0.3cm}

\setcounter{equation}{0}

~~~~Many important discoveries in particle physics (such as flavor mixing in
quarks and in neutrinos) have been made using proton beams with relatively low
energies but high intensities. With this in mind, at a number of laboratories
worldwide interest has been expressed to build facilities that would provide
high-intensity proton beams for various fixed-target experiments
\cite{garoby, baussan}.

Experiments with high-intensity neutrino beams, for instance, are designed
primarily to explore the mass spectrum of the neutrinos and their properties under
the CP symmetry. If there is experimental evidence for CP violation in neutrino
oscillations, it could be used to explain the observed asymmetry between matter
and antimatter in the universe --- a basic precondition for our existence.

Studies of CP violation in the neutrino sector require multi-MW proton beams. Such
beams cannot be provided by any facility that includes proton synchrotrons,
because the beam power in a circular accelerator is limited by the space charge
effects that produce beam instabilities. To increase maximally the beam power of a
`proton driver', it is proposed to build a 15-MW accelerator consisting solely of
a 3-GeV injector linac (PI) and a 17-GeV superconducting ILC-type linac (SCL)
A similar facility was originally proposed in \cite{belusevic}.

At a {\em pulsed} linear accelerator, the {\em beam power} ${\cal P}_{\rm b}^{~}$
increases linearly with the beam energy:
\begin{equation}
{\cal P}_{\rm b}^{~}\,\mbox{[MW]} = {\rm E}_{\rm b}^{~}\,\mbox{[MV]}\times
I\,\mbox{[A]}\times \tau_{\rm p}^{~}\,\mbox{[s]} \times {\cal R}\,\mbox{[Hz]}
\end{equation}
${\rm E}_{\rm b}$ is the {\em beam energy}, $I$ is the {\em average current per
pulse}, $\tau_{\rm p}^{~}$ is the {\em beam pulse length}, and ${\cal R}$ is the
{\em pulse repetition rate}. Assuming ${\rm E}_{\rm b}^{~} = 20\times 10^{3}$ MV,
$I = 32$ mA, $\tau_{\rm p}^{~} = 1.2$ ms and ${\cal R} = 20~{\rm s}^{-1}$,
expression (1) yields  ${\cal P}_{\rm b}^{~} = 15$ MW. This is over an order of
magnitude higher than the beam power at any existing proton synchrotron complex.

\vspace*{0.3cm}
\section{A Two-Linac Proton Driver}
\vspace*{0.3cm}

\begin{figure}[!h]
\vspace{3.3mm}
\begin{center}
\epsfig{file=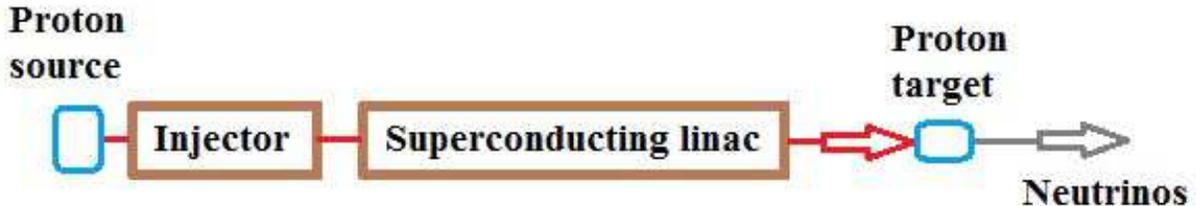,width=0.98\textwidth}
\end{center}
\vskip -5mm
\caption{The layout of the proposed 15-MW `proton driver'. A 3-GeV proton linac (PI) serves as an injector to a superconducting ILC-type linac (SCL), and can also be used to produce, e.g., `cold' neutron beams. Possible prototypes of the PI are
briefly described in the following subsection. Protons accelerated by the SCL to beam energies of 20 GeV are used primarily to create beams for neutrino
oscillation experiments.}
\label{fig:Driver}
\end{figure}

~~~~The layout of the proposed `proton driver' is shown in
Fig.\,\ref{fig:Driver}. A 3-GeV proton linac (PI) serves  as an injector to an
ILC-type superconducting linac (SCL). The main SCL beam parameters are given in
Table 1. Protons accelerated by the SCL to 20 GeV would be used primarily to
create beams for neutrino oscillation experiments.

The typical properties of a 1.3-GHz superconducting ILC-type cavity are presented,
e.g., in \cite{ILCcav}.  Each ILC-type cryomodule for the proposed SCL would
contain eight niobium 9-cell cavities and a focusing quadrupole magnet at its
centre. The inactive regions between cavities or cryomodules (the `packing
fraction') are responsible for a reduction in the average accelerating gradient of
the linac.

The average {\em usable} accelerating gradient in ILC-type cavities is
$\overline{\rm E}_{\rm acc} = 29.3\pm 5.1$ MV/m \cite{reschke}.
Taking into account an estimated linac `packing fraction' of about 70\%, the
effective accelerating gradient of the SCL is E$_{\rm eff} \approx 20$ MV/m.
Hence, the total length of a 17-GeV linac is 850 m.

\begin{table}[t]
\centering
\caption{\bf Parameters of the proposed SCL}
\label{tab:table1}
   \begin{tabular}{ll}
\noalign{\vskip 1mm}   
\hline
\hline
\noalign{\vskip 1mm}
Beam energy ~~&~~ 17 GeV\\
\noalign{\vskip 1mm}
\hline
\noalign{\vskip 1mm}
Effective accelerating gradient ~~&~~ 20 MV/m\\
\noalign{\vskip 1mm}
\hline
\noalign{\vskip 1mm}
Repetition rate ~~&~~ 20 Hz\\
\noalign{\vskip 1mm}
\hline
\noalign{\vskip 1mm}
Protons per pulse ~~&~~ $2.3\times 10^{14}$\\
\noalign{\vskip 1mm}
\hline
\noalign{\vskip 1mm}
Beam pulse length ~~&~~ 1.2 ms\\
\noalign{\vskip 1mm}
\hline
\noalign{\vskip 1mm}
Average current per pulse ~~&~~ 32 mA\\
\noalign{\vskip 1mm}
\hline
\noalign{\vskip 1mm}
Duty factor ~~&~~ 2.4\,\%\\
\noalign{\vskip 1mm}
\hline
\noalign{\vskip 1mm}
RF frequency ~~&~~ 1.3 GHz\\
\noalign{\vskip 1mm}
\hline
\noalign{\vskip 1mm}
Klystron average power ~~&~~ 150 KW\\
\noalign{\vskip 1mm}
\hline
\noalign{\vskip 1mm}
Klystron peak power ~~&~~ 5 MW\\
\noalign{\vskip 1mm}
\hline
\noalign{\vskip 1mm}
Klystron pulse length ~~&~~ 1.5 ms\\
\noalign{\vskip 1mm}
\hline
\noalign{\vskip 1mm}
Peak power per coupler ~~&~~ 312 kW\\
\noalign{\vskip 1mm}
\hline
\hline
   \end{tabular}
\end{table}

Since the length of an ILC 9-cell cavity is 1m, a linac with energy ${\rm E}_
{\rm b} = 17$ GeV would require $N_{\rm cav} = 580$ cavities. The average input rf power per cavity is therefore $\overline{\cal P}_{\rm cav}^{~} \approx 15$ kW,
and the corresponding peak power ${\cal P}_{\rm cav}^{~} \approx 625$ kW. If two
rf couplers per cavity are used, the peak rf power per coupler is reduced to
312 kW. For E$_{\rm acc} \approx 30$ MV/m, ohmic losses in an ILC 9-cell cavity
amount to ${\cal P}_{\rm c}^{~} = 2.4$\,W (plus static loss) in the pulsed mode
with a duty factor $\tau_{\rm p}^{~}\times{\cal R} = 0.024$.

\vspace*{3mm}
For a pulsed ILC-type superconducting linac, one of the currently available rf
sources is the {\em Toshiba} E3736 {\em Multi-Beam Klystron} \cite{yano}. This
source has the following properties: {\em rf frequency} -- 1.3 GHz; {\em peak
rf power} -- 10 MW; {\em average power} -- 150 kW; {\em efficiency} -- 65\%;
{\em pulse length} -- 1.5 ms; {\em repetition rate} -- 10 Hz.

To increase the beam power of the SCL, the pulse repetition rate could be
increased to 20 Hz (see Eq. (1)). The peak power of each klystron would then
have to be reduced to 5 MW in order to maintain its average power at 150 kW for
the klystron pulse length $\tau = 1.5$ ms (see Table 1). In this case a suitable
20 Hz {\em pulse modulator} would be required.

\vspace*{0.3cm}
\subsection{Proton Injector (PI)}
\vspace*{0.3cm}

~~~~A typical proton linear accelerator consists of three main sections:
(1) {\large{\em Front end}}, which includes a proton source and a radiofrequency
quadrupole accelerator; (2) {\large{\em Medium-velocity linac}}, where the proton
beam energy is increased to a few hundred MeV; (3) {\large{\em High-velocity
linac}}, which accelerates protons to energies above 1 GeV.

\begin{figure}[h]
\begin{center}
\epsfig{file=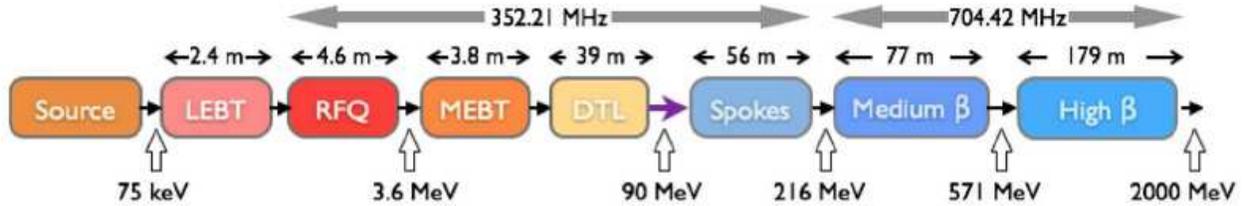,height=0.12\textheight}
\end{center}
\vskip -5mm
\caption{Block diagram of the ESS linac \cite{ESS}. The RFQ and DTL structures
are normal-conducting, while the spoke resonator cavities are superconducting.}
\label{fig:ESS}
\end{figure}

A detailed beam dynamics study of a high-power proton linac is needed in order to
avoid halo formation, a major source of beam loss. Another crucial issue is the
preservation of beam emittance. Either of the following two accelerator designs
could serve as a prototype of the 3-GeV injector linac in Fig.\,\ref{fig:Driver}:

$\bullet$ The {\em European Spallation Source} (ESS) linac, a 2-GeV proton linear
accelerator currently under construction in Lund, Sweden (see
Fig.\,\ref{fig:ESS}). The transverse beam size along the linac varies between
1\,mm and 4\,mm, while the bunch length decreases from 1.2\,cm to 3\,mm towards
the end of the linac. The total length of the accelerator is about 360\,m.

\begin{table}[h]
\centering
\caption{\bf Parameters of the linac described in \cite{Project-X}}
\label{tab:table2}
   \begin{tabular}{ll}
\noalign{\vskip 1mm}
\hline
\hline
\noalign{\vskip 1mm}
Beam kinetic energy ~~~&~~~ 8 GeV\\
\noalign{\vskip 1mm}
\hline
\noalign{\vskip 1mm}
Beam current averaged over the pulse ~~~&~~~ 25 mA\\
\noalign{\vskip 1mm}
\hline
\noalign{\vskip 1mm}
Pulse repetition rate ~~~&~~~ 10 Hz\\
\noalign{\vskip 1mm}
\hline
\noalign{\vskip 1mm}
Pulse length ~~~&~~~ 1 ms\\
\noalign{\vskip 1mm}
\hline
\noalign{\vskip 1mm}
Beam pulsed power ~~~&~~~ 200 MW\\
\noalign{\vskip 1mm}
\hline
\noalign{\vskip 1mm}
Beam average power ~~~&~~~ 2 MW\\
\noalign{\vskip 1mm}
\hline
\noalign{\vskip 1mm}
Wall-plug power (estimate) ~~~&~~~ 12.5 MW\\
\noalign{\vskip 1mm}
\hline
\noalign{\vskip 1mm}
Total length ~~~&~~~ 678 m\\
\noalign{\vskip 1mm}
\hline
\hline
   \end{tabular}
\end{table}

\vspace*{3mm}
$\bullet$ An 8-GeV linac for a `proton driver' at Fermilab \cite{Project-X}. The
linac consits of a front-end section that accelerates protons to energies of about
420 MeV, and a high-energy section operating at rf frequencies of 325 MHz and 1300
MHz respectively. A room temperature radio frequency quadrupole (RFQ) and short
H-type resonators are proposed for the initial acceleration of proton beams to
energies of $\sim\,10$ MeV. In the energy range 10--420 MeV the acceleration is
provided by superconducting spoke resonators, and in the high-energy section by
ILC-type elliptical cell cavities.

\begin{figure}[t]
\begin{center}
\epsfig{file=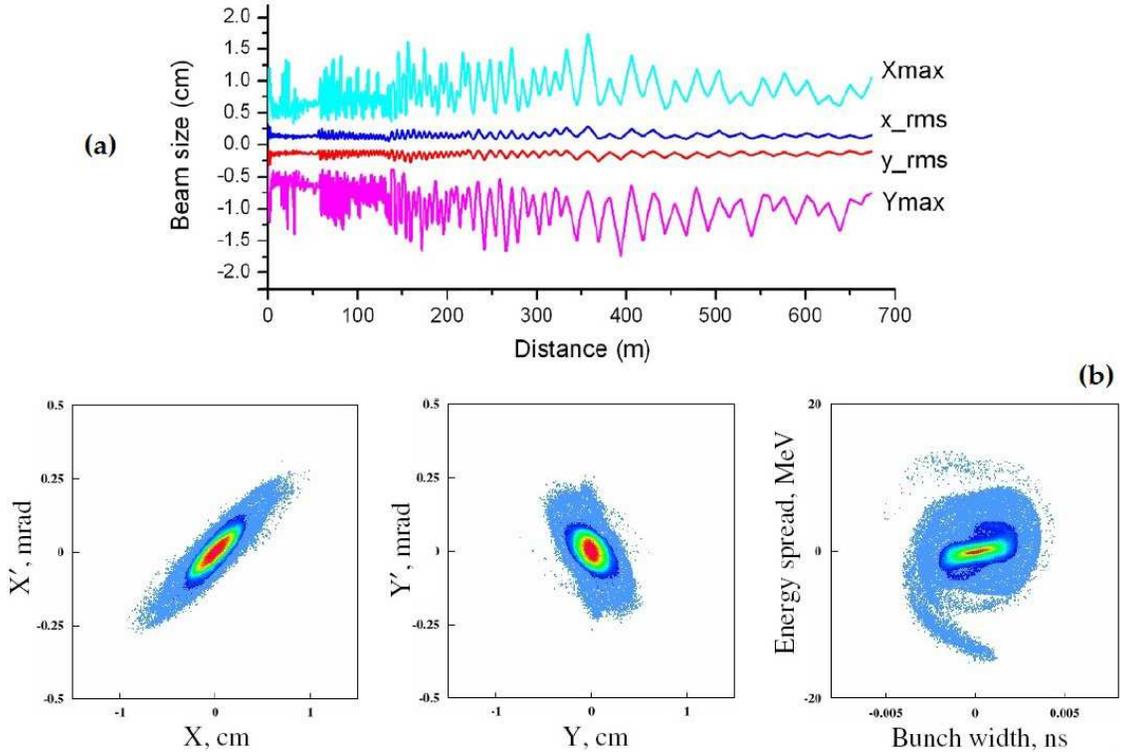,width=0.9\textwidth}
\end{center}
\vskip -5mm
\caption{(a) Transverse envelopes of a 43.25 mA beam along the 8-GeV proton linac
described in reference \cite{Project-X}; (b) Phase-space plots of the beam at the
end of the linac. Credit: P.\,N. Ostroumov.}
\label{fig:Ostroumov}
\end{figure}

The basic parameters of the proposed 8-GeV linac are listed in Table 2. The beam
physics and the lattice design of the linac are described in \cite{Project-X}.
Some of the results of end-to-end beam dynamics simulations are presented in
Fig.\,\ref{fig:Ostroumov}  (Section 7 in \cite{Project-X}). The block diagram
of the linac is shown in Subsection 3.3 of \cite{Project-X}.

\vspace*{0.3cm}
\subsection{Proton Target and Magnetic Horn for Neutrino Beams}
\vspace*{0.3cm}

~~~~When designing a multi-MW neutrino beam facility, one of the main challenges
is to build a {\em proton target} that could withstand the strong pressure waves
created by short beam pulses, dissipate large amounts of deposited energy,  and
survive long-term effects of radiation damage. Computer simulations of the pion
production and energy deposition in different targets (solid tungsten bars; jets
of liquid mercury, liquid gallium and tungsten powder) are described in \cite{back}.

A 4\,MW target station comprising a liquid mercury jet inside a 20\,T solenoidal magnetic field was tested in 2007 at CERN \cite{mcdonald}. An alternative design
is a rotating, gas-cooled tungsten target that would require the least amount of
development effort, and would have good thermal and mechanical properties
\cite{ESS}. To reduce the beam power delivered to a target, a 15-MW proton beam
could be separated by a series of magnets into three beam lines. Each of the
5-MW beams would be guided to an assembly consisting of three targets and the same
number of {\em magnetic horns}, devices that focus the charged particles produced
in the proton target (see, e.g., \cite{baussan1}).

For proton beam pulses lasting 1 ms, a DC horn has been designed at KEK by
Yukihide Kamiya \cite{kamiya}. The toroidal magnetic field of the horn, generated
by hollow aluminium conductors that contain cooling water, has the same strength
at all radii: B$(r)=$ 0.2 T. The outer radius of the magnet, $R$, is determined by
\begin{equation}
R = L\tan (\theta ) + \ell\tan (\theta /2)
\end{equation}
where $\theta\approx 0.03 + 0.3/p$ is the initial angle of a charged pion with
respect to the proton beam direction, $L$ is the distance from the target to the horn, $\ell = 5$\,m is the length of the magnet, and $p$ is the pion momentum.
The total power generated in the conductors is about 10 MW.

\vspace*{0.3cm}
\section{Significance of Neutrino Oscillations}
\vspace*{0.3cm}

~~~~Observation of the quantum-mechanical phenomenon of neutrino oscillations
implies that at least one kind of neutrino has non-zero mass. The actual mechanism of neutrino mass generation is still not known. In the {\em Standard Model} of
particle physics, fermions acquire mass through interactions with the Higgs field,
an entity that permeates the universe. These interactions involve both left- and
right-handed versions of each fermion. However, only left-handed neutrinos (and
righ-handed antineutrinos) have been observed so far. The discovery of neutrino
oscillations in 1998 represents, therefore, compelling experimental evidence for
the incompleteness of the Standard Model as a description of nature.

The universe contains about a billion neutrinos for every quark or electron.
Relic neutrinos from the early universe are almost as abundant as cosmic microwave
background photons (about 330 neutrinos and antineutrinos of all species per
cm$^{3}$, compared to about 410 photons per cm$^{3}$). Cosmological data suggest
that the combined mass of all three neutrino species (`flavors') is at least a
million times smaller than that of the next-lightest particle, the electron.

The phenomenon of neutrino oscillations implies not only the existence of
neutrino mass, but also of {\em neutrino mixing}: the neutrinos of definite flavor
are not particles of definite mass (mass eigenstates), but coherent
quantum-mechanical superpositions of such states. Converesely, each neutrino of
definite mass is a superposition of neutrinos of definite flavor.

A neutrino state with a well-defined flavor, $|\nu_{\alpha}^{~}\rangle$, can be
expressed in terms of mass eigenstates $|\nu_{i}^{~}\rangle$ as follows:
\begin{equation}
|\nu_{\alpha}^{~}\rangle = \sum_{i\,=\,1}^{3}U^{*}_{\alpha i}|\nu_{i}^{~}\rangle ,
\hspace*{2cm}\alpha = e,~\mu~\mbox{or}~\tau
\end{equation}
Here $U$ is a complex unitary matrix, called the {\em mixing matrix}. For
neutrino oscillations in vacuo, one can use this expression to derive the
probability that a different flavor eigenstate $|\nu_{\beta}^{~}\rangle$ will be
observed at time $t$:
\begin{equation}
{\cal P}(\nu_{\alpha}^{~}\rightarrow\nu_{\beta}^{~}) = \delta_{\alpha\beta}^{~} -
4\sum_{i>j}{\rm Re}\,(U^{*}_{\alpha i}U^{~}_{\beta i}U^{~}_{\alpha j}U^{*}_
{\beta j})\sin^{2}\phi + 2{\rm Im}\,(U^{*}_{\alpha i}U^{~}_{\beta i}U^{~}_
{\alpha j}U^{*}_{\beta j})\sin 2\phi
\end{equation}
In Eq. (4),
\begin{equation}
\phi \equiv \frac{\Delta m_{ij}^{2}L}{4{\rm E}}
\end{equation}
where $\Delta m_{ij}^{2} = m_{i}^{2}-m_{j}^{2}$ are the mass-squared differences,
E is the common energy of all $|\nu_{i}^{~}\rangle$ components, and $L$ is the
distance travelled by a flavor eigenstate $|\nu_{\alpha}^{~}\rangle$ to the
detector where $|\nu_{\beta}^{~} \rangle$ is observed. Here it is assumed that
$|\nu_{\alpha}^{~}\rangle$, produced at a neutrino source, propagates as a
superposition of mass eigenstates $|\nu_{i}^{~}\rangle$.

The neutrino oscillation rate depends, in part, on ($a$) the difference between
neutrino masses and ($b$) the three parameters in the mixing matrix known as the
{\em mixing angles}. As one can infer from Eq. (4), ${\cal P}(\nu_ {\alpha}^{~} \rightarrow\nu_{\beta}^{~}) = \delta_{\alpha\beta}^{~}$ if all $\Delta m^{2} = 0$.
Thus, observation of neutrino oscillations implies that at least one kind of
neutrino has non-zero mass, as already stated.

The complex phase factors in the mixing matrix are associated with the violation
of CP symmetry. In neutrino oscillations, CP violation is observed only if all the
mixing angles and all the mass differences are nonvanishing \cite{bellini}. If there is no CP violation, which means that ${\cal P}(\nu_{\alpha}^{~}\rightarrow \nu_{\beta}^{~}) = {\cal P}(\bar{\nu}_ {\alpha}^{~}\rightarrow\bar{\nu}_{\beta} ^{~})$, the last term in Eq. (4) vanishes.

As mentioned in the Introduction, experimental evidence for CP violation in
neutrino oscillations could be used to explain the observed asymmetry between
matter and antimatter in the universe. Since this asymmetry is a basic
precondition for our existence, search for CP violation in the lepton sector is
the holy grail of neutrino physics.

\vspace*{0.3cm}
\section{Acknowledgements}
\vspace*{0.3cm}

~~~~For valuable comments and suggestions concerning various aspects of this work
I am grateful to Y. Kamiya and K. Oide. I wish to express my special gratitude to
Kaoru Yokoya for his help and encouragement.

This paper is based on the idea originally presented in the KEK Preprint 2014-35
(2014), which was subsequently published as an arXiv e-print (arXiv:1411.4874) and
\cite{belusevic}.




\begin{thebibliography}{99}
\vspace*{0.5cm}

\bibitem{garoby}
R. Garoby et al., \textit{Proton drivers for neutrino beams and other high
intensity applications}, Journal Phys. Conf. Series 408, 012016 (2013).

\bibitem{baussan}
E. Baussan et al., {\em A very intense neutrino super beam experiment for
leptonic} CP {\em violation discovery based on the European spallation source
linac}, Nucl. Phys. B 885, 127--149 (2014).

\bibitem{belusevic}
R. Belusevic, {\em A multi-MW proton/electron facility at} KEK, J. Appl. Math.
Phys. 5, 1222--1242 (2017).

\bibitem{ILCcav}
A. Yamamoto, {\em Superconducting RF cavity development for the International
Linear Collider}, IEEE Trans. Appl. Supercond. Vol. 19, No. 3, pp 1387--1393
(2009); T. Saeki et al., {\em Study on fabrication of superconducting RF 9-cell
cavity for}  ILC {\em at} KEK, Proc. IPAC 2013, Shanghai, China, pp 3132--3134
(2013).

\bibitem{reschke}
D. Reschke, {\em Infrastructure, methods and test results for the testing of} 800
{\em series cavities for the European} XFEL, Proc. SRF 2013, Paris, France,
pp 812--815 (2013).

\bibitem{yano}
A. Yano et al., \textit{The Toshiba} E3736 {\em Multi-Beam Klystron},
Proc. LINAC 2004, L\"{u}beck, Germany, pp 706--708 (2004).

\bibitem{ESS}
H. Danared, M. Lindroos and C. Theroine, ESS: \textit{neutron beams at the
high-intensity frontier}, CERN Courier, June 2014, pp 21--24.

\bibitem{Project-X}
P.\,N. Ostroumov, {\em Physics design of the} 8 GeV H-{\em minus linac},
New J. Phys. 8 (2006) 281.

\bibitem{back}
J. Back et al., {\em Particle production and energy deposition studies for the
neutrino factory target station}, Phys. Rev. Spec. Topics --- Accelerators and
Beams 16, 021001 (2013).

\bibitem{mcdonald}
K.\,T. McDonald et al., {\em The} MERIT {\em high-power target experiment at the}
CERN PS, in Proc. First International Particle Accelerator Conf., Kyoto, Japan,
pp 3527--3529 (2010).

\bibitem{baussan1}
E. Baussan et al., {\em Target, magnetic horn and safety studies for the} CERN
{\em to Frejus Super Beam}, Journal of Phys. Conf. Series 408, 012061 (2013).

\bibitem{kamiya}
Y. Kamiya, {\em DC neutrino horn}; the manuscript (in Japanese) has not been
published.

\bibitem{bellini}
G. Bellini et al., {\em Neutrino Oscillations}, Adv. High. En. Phys., Volume 2014,
Article ID 191960 (2014).


\end{thebibliography}
\end{document}